\documentclass[aps,twocolumn,showpacs]{revtex4}
\usepackage{amsmath}
\usepackage{amsfonts}
\usepackage{graphicx}

\begin{document}

\title{Two-body scattering in a trap and a special periodic phenomenon
sensitive to the interaction}
\author{Y. Z. He}
\author{C. G. Bao}
\thanks{Corresponding author: stsbcg@mail.sysu.edu.cn}
\affiliation{State Key Laboratory of Optoelectronic Materials and Technologies, Sun
Yat-Sen University, Guangzhou, 510275, P.R. China}

\begin{abstract}
Two-body scattering of neutral particles in a trap is studied theoretically.
The control of the initial state is realized by using optical traps. The
collisions inside the trap occur repeatedly; thereby the effect of
interaction can be accumulated. Two periodic phenomena with a shorter and a
much longer period, respectively, are found. The latter is sensitive to the
interaction. Instead of measuring the differential cross section as usually
does, the measurement of the longer period and the details of the periodic
behavior might be a valid source of information on weak interactions among
neutral particles.
\end{abstract}

\maketitle

\date{\today}

The determination of interactions among microscopic particles is an
important topic in physics. Historically, the measurement of the
cross sections of 2-body scattering provides an important source of
information. For neutral atoms and molecules the determination is
more difficult because the interaction is in general weak and the
initial momentum of the incident particle is difficult to be
precisely controlled. However, if the scattering occurs in a trap,
new phenomena previously unknown might emerge. Due to the recent
progress in the techniques of trapping atoms (molecules) by using
optical traps \cite{r_SJ1998,r_BMD2001,r_WP2009}, trapped scattering
might be eventually experimentally realized. In a previous paper a
model was proposed to study the trapped 2-body scattering
theoretically \cite{r_LZB2009}. Instead of a single collision,
numerous repeated collisions have been found. Thereby the effect of
interaction can be accumulated and enlarged, and might be eventually
detected. This favors the determination of very weak interaction.
The emphasis of that paper is placed on the study of the spin-flip
phenomenon and the determination of the parameter $g_0$ (the
strength of the channel with total spin zero) of the $^{52}Cr$
atoms. This paper is also dedicated to trapped scatterings; however
the emphasis is placed on the study of a specific periodicity
emerging from the repeated collisions. It turns out that the
associated period is sensitive to the interaction. Therefore, in
addition to the measurement of cross sections, the observation of
the special periodic phenomenon might be also a valid way for the
determination of interaction. The model, related theoretical
derivation, and numerical results are given below.

We propose a device containing two deep optical traps initially. One
is close to the origin, while the other one is far away. Each trap
provides a harmonic potential. Thus the total potential was
$U_p(\mathbf{r})=\frac{1}{2}\mathcal{M}\omega_p^2(|\mathbf{r-a}|^2+|\mathbf{r-b}|^2)$,
where $\mathbf{a}$ and $\mathbf{b}$ are two given vector (norms
$a\gg b$), and $\mathcal{M}$ is the mass of the particle involved.
Each trap contains a particle in the lowest harmonic oscillator
(h.o.) state. The two particles are assumed to be identical bosons
(the generalization to fermions is straight forward), the
interaction is assumed to be spin-independent, and $\omega_p$ is
large enough so that the particles are well localized initially and
the overlap of their wave functions is negligible. Suddenly the two
deep traps are cancelled. Instead, a broader new trap located at the
origin $U_{evol}(r)=\frac{1}{2}\mathcal{M}\omega^2 r^2$ is created,
$\omega <\omega_p$. Since the initial state is not an eigenstate of
the new Hamiltonian, the system begins to evolve. The evolution is
affected not only by $U_{evol}(r)$ but also by the interaction
$V(|\mathbf{r}_i\mathbf{-r}_j|)$. In what follows the details of the
evolution is studied, two-body collisions occurring repeatedly are
found, and the effect of interaction is demonstrated.

Let $\hbar\omega$ and $\sqrt{\hbar /\mathcal{M}\omega}$ be used as
units of energy and length. The normalized initial state is
\begin{equation}
 \Psi _I
 =\frac{1}{\sqrt{2}}
  (1+P_{1,2})
  (\frac{\eta }{\pi})^\frac{6}{4}
  \exp[-\frac{\eta}{2}(|\mathbf{r}_1\mathbf{-a}|^2
       +|\mathbf{r}_2\mathbf{-b}|^2)]
 \label{e01_PsiI}
\end{equation}
where $P_{1,2}$ implies an interchange of $1$ and $2$, and
$\eta=\omega_p/\omega$. We consider the case that $\mathbf{a}$ is
lying along the $X-$axis, while $\mathbf{b}$ along the negative
$Z-$axis. When $\mathbf{R}=(\mathbf{r}_1+\mathbf{r}_2)/2$ and
$\mathbf{r}=\mathbf{r}_2-\mathbf{r}_1$ for the c.m. and relative
motions, respectively, are introduced, and the h.o. states of the
new trap are selected as base functions, the initial state can be
expanded as
\begin{equation}
 \Psi_I
 =\sum_{NLnlJM}
  B_{NLnl}^{JM}
  [\varphi_{NL}^{(2)}(\mathbf{R})\
   \varphi_{nl}^{(1/2)}(\mathbf{r})]_{JM}
 \label{e02_PsiI}
\end{equation}
where $\varphi _{nl}^{(\mu )}(\mathbf{r})$ is a normalized
eigenstates of the Hamiltonian $-\frac{1}{2\mu}\nabla
_{\mathbf{r}}^2+\frac{1}{2}\mu r^2$ with the eigenenergy $2n+l+3/2$.
$L$ and $l$ are coupled to the total orbital angular momentum $J$
and $M$.
\begin{eqnarray}
 B_{NLnl}^{JM}
 &=&\sqrt{2}
  (\frac{\eta}{\pi})^{\frac{3}{2}}
  \sum_{m}
  C_{L,M-m,l,m}^{J,M}  \nonumber \\
 && \times
  \{\int d\mathbf{R}\
    \varphi_{N,L,M-m}^{(2)}(\mathbf{R})  \nonumber \\
 && \times\ \ \
  e^{-\frac{\eta}{2}
     [2R^2+a^2-2R(a\sin\theta_R \cos\phi_R-b\cos\theta_R)]}\}  \nonumber \\
 &&\times
  \{\int d\mathbf{r}\
    \varphi_{nlm}^{(1/2)}(\mathbf{r})  \nonumber \\
 &&\times\ \ \
  e^{-\frac{\eta }{2}
     [\frac{1}{2}r^2+b^2+r(a\sin\theta_r\cos\phi_r +b\cos\theta_r)]}\}
 \label{e03_B}
\end{eqnarray}
where the Clebsch-Gordan coefficients are introduced, $\theta_R$ and
$\phi_R$ are the spherical polar coordinates, and so on, $l$ must be
even for boson systems.

The new Hamiltonian in terms of $\mathbf{R}$ and $\mathbf{r}$ governing the
evolution is
\begin{equation}
 \left\{
 \begin{array}{rll}
  H_{evol} & = & H_R+H_r \\
  H_R & = & -\frac{1}{4}\nabla _{\mathbf{R}}^2+R^2 \\
  H_r & = & -\nabla _{\mathbf{r}}^2+\frac{1}{4}r^2+V(r)
 \end{array}
 \right.
 \label{e04_H}
\end{equation}
Using $\varphi _{nl}^{(1/2)}$ as base functions, the eigenstates of
$H_r$ can be obtained via a diagonalization and can be expanded as
\begin{equation}
\psi_{l,i}(\mathbf{r}) =\sum_n D_n^{l,i} \varphi_{nl}^{(1/2)}(\mathbf{r})
\label{e05_psili}
\end{equation}
where $i$ denotes the $i-$th eigenstate of the $l-$series in the
order of increasing energy. The associated energy is denoted by
$E_{l,i}$. Reversely, $\varphi _{nl}^{(1/2)}$ can also be expanded
in terms of $\psi _{l,i}$. Then, starting from $\Psi_I$, the
time-dependent solution of $H_{evol}$ is
\begin{eqnarray}
 \Psi (\tau )
 &=&e^{-iH_{evol}\ \tau }
  \Psi_I  \nonumber \\
 &=&\sum_{NLnlJM}
  B_{NLnl}^{JM}
  \sum_{n'i}
  D_n^{l,i}
  D_{n'}^{l,i}\
  e^{-i(2N+L+3/2+E_{l,i})\tau}  \nonumber \\
 && \times
  [\varphi_{NL}^{(2)}(\mathbf{R})\
   \varphi_{n'l}^{(1/2)}(\mathbf{r})]_{JM}
 \label{e06_Psitau}
\end{eqnarray}
where $\tau =\omega t$. In principle, the above solution is exact only if
the summation covers infinite terms. However, when the interaction is not
strong, qualitatively accurate solutions can be obtained if the number of
base functions is large enough. This is shown below.

We define the time-dependent one-body density from $\Psi (\tau )$ as
\begin{equation}
 \rho(\mathbf{r}_1,\tau)
 \equiv \int
  d\mathbf{r}_2\
  \Psi^*(\tau)
  \Psi(\tau)
 \label{e07_rho}
\end{equation}
In order to obtain $\rho$, the Talmi-Moshinsky coefficients relating two
sets of coordinates are introduced as
\begin{eqnarray}
 &&[\varphi _{NL}^{(2)}(\mathbf{R})\
  \varphi_{n'l}^{(1/2)}(\mathbf{r})]_{JM}  \nonumber \\
 &&=\sum_{n_1 l_1 n_2 l_2}
  a_{n_1 l_1 n_2 l_2}^{NLn'l,J}
 [\varphi_{n_1 l_1}^{(1)}(\mathbf{r}_1)\
  \varphi_{n_2 l_2}^{(1)}(\mathbf{r}_2)]_{JM}
 \label{e08_varphi}
\end{eqnarray}
The analytical form of the coefficients can be found in
\cite{r_WT1981,r_BM1966,r_BTA1960}. From Eqs.~(\ref{e06_Psitau}) and
(\ref{e08_varphi}), making use of the orthonormality of the base
functions, the integration in Eq.~(\ref{e07_rho}) is easy to carry
out, and it is straight forward to obtain the analytical expression
of $\rho(\mathbf{r}_{1},\tau)$. Incidentally, since the wave
function is symmetrized, the behaviors of the two particles are
exactly the same. The observation of only one particle is
sufficient.

To obtain numerical results as examples, it is first assumed that $a=2$, $b=0
$, $\eta =2$, and the interaction contains a stronger repulsive core and a
weaker attractive tail as
\begin{equation}
 V(r)
 =\left\{
  \begin{array}{rll}
   & V_0 , & \mbox{ if } r<0.2 \\
   & -C_6/r^6 , & \mbox{ else }
  \end{array}
  \right.
 \label{e09_V}
\end{equation}
where $V_0$ and $C_6$ are positive numbers, and $V_0 \gg C_6$. We
define $K=2(N+n)+L+l$ to control the dimension of the base. Mostly,
the base functions with $K\leq K_{\max}=20$ are adopted in the
following calculation.
\begin{figure}[htbp]
 \centering
 \resizebox{0.95\columnwidth}{!}{\includegraphics{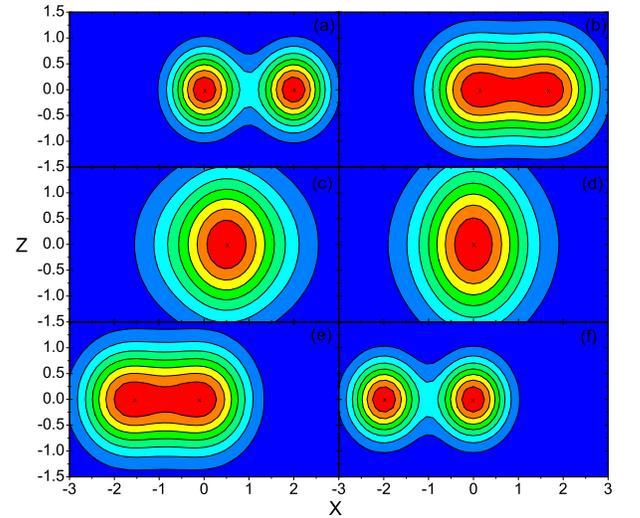}}
 \caption{(Color online) $\protect\rho(\mathbf{r}_1,\protect\tau)$
plotted on the $X-Z$ plane. $\protect\tau=0$, $\protect\pi/6$,
$\protect\pi/3$, $\protect\pi/2$, $5\protect\pi/6$ and
$\protect\pi$, respectively, for (a) to (f). The parameters are
$\protect\eta=2$, $a=2$, $b=0$, $V_0=10$, $r_0=0.2$ and $c_6=-0.3
r_0^6$. The units of energy and length in this paper are
$\hbar\protect\omega$ and $\protect\sqrt{\hbar
/\mathcal{M}\protect\omega}$, where $\protect\omega$ has not yet
been specified. Every maximum in the panels is marked by a cross.
The values associated with the inmost contours of (a) to (f) are
$0.224$, $0.098$, $0.073$, $0.063$, $0.098$ and $0.224$. Thus the
peaks in (a) and (f) are much higher. Therefore, the particles are
better localized when they separate from each other. The values of
the outmost contours are from $0.01$ to $0.03$.}
 \label{fig1}
\end{figure}

When $V_0=10$, $C_6=-0.3$, and $\tau$ is given at a number of values
in the early stage of evolution, $\rho(\mathbf{r}_1,\tau)$ plotted
on the $X-Z$ plane is shown in Fig.~\ref{fig1}. \ref{fig1}a is for
the initial case. When the evolution begins, the outside particle
moves toward the center and collides with the target particle as
shown in \ref{fig1}b and \ref{fig1}c. When $\tau =\pi /2$ both
particles are close to the center as in \ref{fig1}d. Afterward a
particle begins to leave as in \ref{fig1}e, and will arrive at the
opposite end at $\tau=\pi$ as in \ref{fig1}f. Then the process
repeats but in reverse direction. When $\tau=2\pi$ the system
recovers its initial status. This is a kind of periodic behavior
originating from the harmonic trap. If the interaction is neglected,
the recovery would be exact. However, due to the interaction,
$\rho(\mathbf{r}_1,\tau)$ is not exactly equal to
$\rho(\mathbf{r}_1,\tau +2\pi)$. With the above parameters, the
deviation is very small. Say, when $\mathbf{r}_1=\mathbf{a}$, we
have $\rho(\mathbf{a},0)=0.2553$ and $\rho(\mathbf{a},2\pi)=0.2544$.
Obviously, in the period $(0,2\pi)$ the system undergoes a pair of
collisions. When the time goes on, a series of head-on collisions
occur repeatedly in the trap. Although the effect of interaction on
a round of collision is weak, the effect of many rounds might be
accumulated and therefore might become strong. This is shown below.
\begin{figure}[htbp]
 \centering
 \resizebox{0.95\columnwidth}{!}{\includegraphics{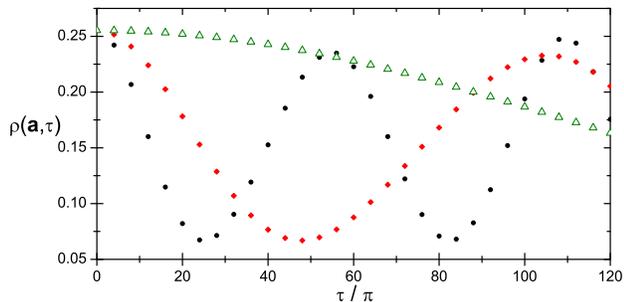}}
 \caption{(Color online) $\protect\rho(\mathbf{a},\protect\tau)$
plotted against $\protect\tau$, where $\mathbf{a}$ denotes the
initial position of the incident particle and $\protect\tau$ is
given only in discrete values $2k\protect\pi$, where $k$ is an
integer from $0$ are $60$. Three choices of $V_0$, namely, $20$,
$10$, and $2$ are adopted, and the associated $\protect\rho$ are
marked with circles, diamonds, and triangles, respectively (however
the values of $\protect\rho$ associated with an odd $k$ have been
neglected just for simplicity). The other parameters are the same as
in Fig.~\protect\ref{fig1}.}
 \label{fig2}
\end{figure}

In Fig.~\ref{fig2} $\rho(\mathbf{a},\tau)$ is given at $\tau=2k\pi$,
where $k $ is an integer. It implies that $\rho $ is observed at the
initial position of the incident particle repeatedly. If the
interaction is removed, all the symbols in the figure would lie
along a horizontal line (implying an exact periodicity with a period
$2\pi$). However, the interaction causes a deviation. The deviation
would become larger if the interaction is stronger (the black
circles to be compared with the triangles) and/or if $\tau$ is
larger. It is found that, when the time goes on,
$\rho(\mathbf{a},2k\pi)$ against increasing $k$ will first arrive at
a minimum, then arrive at the second maximum which is a little lower
than the first maximum at $\tau=0$, then again a minimum, and
afterward arrive at the third maximum with a height close to the
first maximum. This behavior will repeat again and again. E.g., for
$V_0=20$ (black circles), the first, second and third maxima appear
at $\tau=0$, $54\pi$, and $110\pi$, respectively. Whereas for
$V_0=10$ (red diamonds), they appear at $0$, $104\pi$ and $210\pi$.

On the other hand, making use of the expansion
Eq.~(\ref{e06_Psitau}), we calculate the overlap
$|\langle\Psi(0)|\Psi(2k\pi)\rangle|$ which varies with $k$. When
$k$ leads to a minimum (maximum) of $\rho(\mathbf{a},2k\pi)$, the
overlap is small (close to one). For examples, with the parameters
for Fig.~\ref{fig1}, when $\tau=48\pi$, $104\pi$ and $210\pi$
associated with the first minimum, the second maximum, and the third
maximum, respectively, we have $|\langle\Psi(0)|\Psi(\tau
)\rangle|=0.103$, $0.918$ and $0.994$. It is further noted that the
imaginary part of $\langle\Psi(0)|\Psi(210\pi)\rangle$ is very
small. Thus, $\Psi(210\pi)$ is extremely close to the initial state.
Therefore, from the time-dependent Schr\"{o}dinger equation, we know
that what happens during the interval $(0,210\pi)$ will nearly
exactly repeat again in the next interval $(210\pi,420\pi)$, and so
on. It implies the existence of another nearly periodic behavior.
Thus, there are two distinct periodic behaviors. One has a period
$2\pi$, and the other one has a much longer period (say, for the
above case, the period is $210\pi$).
\begin{figure}[htbp]
 \centering
 \resizebox{0.95\columnwidth}{!}{\includegraphics{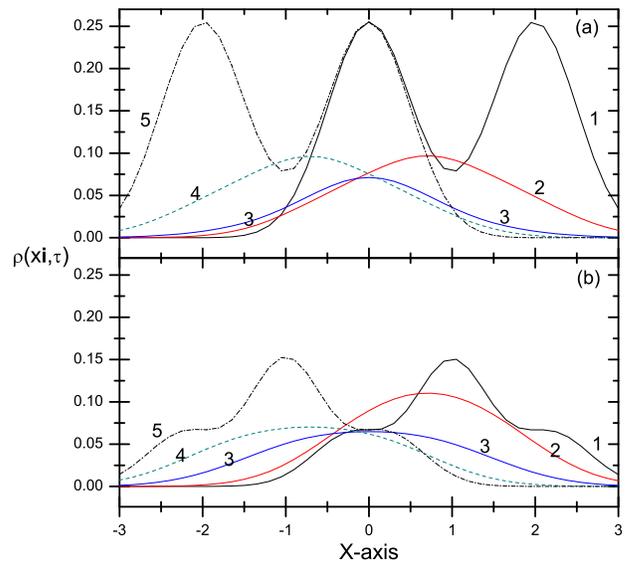}}
 \caption{(Color online) $\protect\rho(\mathbf{r}_1,\protect\tau)$
plotted along the $X-$axis. The curves "1" to "5" in
\protect\ref{fig3}a have $\protect\tau$ from $0$ to $\protect\pi$
with a step $\protect\pi/4$. Those of \protect\ref{fig3}b have
$\protect\tau$ from $48\protect\pi$ to $49\protect\pi$ with the same
step. The parameters are the same as in Fig.~\protect\ref{fig1}.}
 \label{fig3}
\end{figure}

\begin{figure}[htbp]
 \centering
 \resizebox{0.95\columnwidth}{!}{\includegraphics{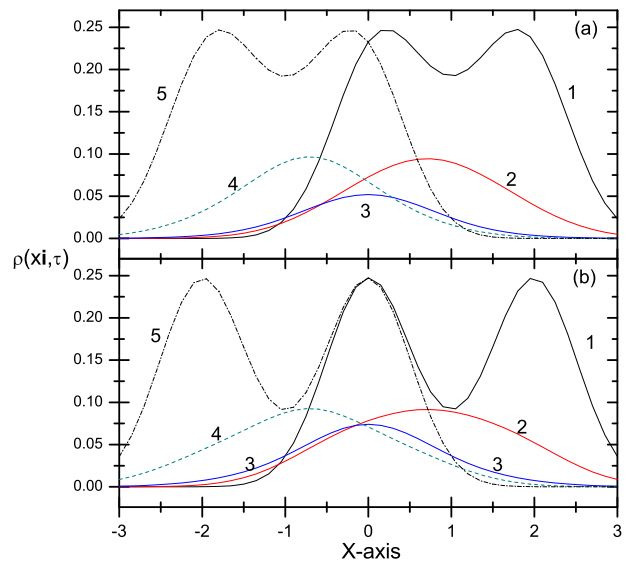}}
 \caption{(Color online) The same as Fig.~\protect\ref{fig3}, but the
domain of $\protect\tau$ is $(104\protect\pi,105\protect\pi)$ in
\protect\ref{fig4}a and $(210\protect\pi,211\protect\pi)$ in
\protect\ref{fig4}b.}
 \label{fig4}
\end{figure}

The periodic behaviors can be shown in more detail by observing
directly the densities. Let $\mathbf{r}_1=x\mathbf{i}$, where
$\mathbf{i}$ is a unit vector along the $X-$axis. The distribution
of $\rho(x\mathbf{i},\tau)$ along the $X-$axis is plotted in
Figs.~\ref{fig3} and \ref{fig4}, where $\tau $ is given at a number
of values. Fig.~\ref{fig3}a describes the evolution in the interval
$(0,\pi)$, where the curve "1" is for the initial state. From "1" to
"5" $\tau$ goes from $0$ to $\pi$ with a step $\pi/4$. One can see
how the two particles undergo a round of collision. In fact,
Fig.~\ref{fig1} and Fig.~\ref{fig3}a describe the same thing except
that $\rho$ is plotted on the $X-Z$ plane in the former but only
along the $X-$axis in the latter. When $\tau$ goes from $\pi$ to
$2\pi$, the process occurring in $(0,\pi)$ will repeat again but in
reverse direction. Thereby the cycle with the $2\pi$ period is
completed. Fig.~\ref{fig3}b describes the evolution in the interval
$(48\pi ,49\pi)$ associated with the first minimum of the diamonds
in Fig.~\ref{fig2}. The peaks in \ref{fig3}b are much lower than
those of \ref{fig3}a, and these peaks are located at different
places. Therefore the system behaves differently in the two
intervals, and the previous clear picture of a head-on collision
becomes ambiguous. Fig.~\ref{fig4}a is associated with the second
maximum. The collision can be roughly seen but is not as clear as in
Fig.~\ref{fig3}a. Fig.~\ref{fig4}b is associated with the third
maximum, and is nearly identical to \ref{fig3}a. Therefore the cycle
with the longer period is completed.
\begin{figure}[htbp]
 \centering
 \resizebox{0.95\columnwidth}{!}{\includegraphics{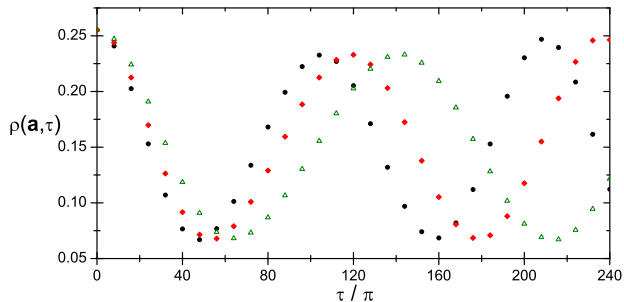}}
 \caption{(Color online) Similar to Fig.~\protect\ref{fig2}, but
$V_0$ is fixed at $10$ and $c_6$ has three choices: $-0.3 r_0^6$,
$-1.5 r_0^6$ and $-3 r_0^6$. The values of
$\protect\rho(\mathbf{a},\protect\tau)$ are, respectively, marked by
black circles, red diamonds, and triangles. The other parameters are
the same as in Fig.~\protect\ref{fig1}. This figure has a longer
range than that in Fig.~\protect\ref{fig2}, and only the values of
$\protect\rho$ with $\protect\tau=8k\protect\pi$ are shown.}
 \label{fig5}
\end{figure}

In Fig.~\ref{fig5}, the strength of the attractive tail has been given at
three values. In each case a slightly lower peak followed by a higher peak
(the third maximum) appears again. With the three choices of $c_6$, the
third maximum appears at $\tau=210\pi$, $236\pi$, and $282\pi$,
respectively. At these three instants, $|\langle\Psi(0)|\Psi(\tau)\rangle|$
are all equal to $0.994$ and the associated imaginary parts are very small.
It implies a nearly exact recovery. Thus the nearly periodic behavior with
the much longer period appears again. When $b\neq 0$ (i.e., the target
particle is not at the center initially), the above qualitative features
remain. In particular, for a specific interaction, the longer period is not
changed with $b$. Thus we conclude that the periodic phenomenon is common to
trapping 2-body scatterings. It is emphasized that this phenomenon is
sensitive to the interactions. A stronger repulsive (attractive) force would
lead to a shorter (longer) long-period.

The accuracy of the above numerical results depends on $K_{\max}$.
As an example selected values of $\rho(\mathbf{a},\tau)$ are listed
in Tab.~\ref{revtab1} to show the dependence.
\begin{table}[tbph]
 \caption{$\protect\rho(\mathbf{a},\protect\tau)$ with three choices
of $K_{\max}$. The parameters involved are the same as those for
Fig.~\protect \ref{fig1}.}
 \label{revtab1}
 \begin{ruledtabular}
  \begin{tabular}{lccc}
  $K_{\max }$                & 12    & 16    & 20      \\
  \hline
  $\rho (\mathbf{a},0)$      & 0.253 & 0.255 & 0.255   \\
  $\rho (\mathbf{a},50\pi)$  & 0.067 & 0.068 & 0.068   \\
  $\rho (\mathbf{a},100\pi)$ & 0.229 & 0.230 & 0.229
  \end{tabular}
 \end{ruledtabular}
\end{table}

The convergency appears to be satisfying. Thus the numerical results
obtained by using $K_{\max}=20$ are accurate enough in qualitative sense.

In this paper the traditional 2-body scattering is considered under a new
environment, namely, in a trap. Due to the trap, the two particles collide
with each other repeatedly. Therefore, even the interaction is weak; the
effect of interaction can be accumulated via the repeated collisions.
Besides, comparing with the case of charged particles, the initial
scattering states of neutral particles are more difficult to control. This
disadvantage can be overcome by using optical traps. Furthermore, two
periodic phenomena are found. They are essentially caused by the trap and by
the interaction, respectively. The period of the latter is much longer and
is sensitive to the interaction. Instead of measuring the differential cross
section as usually does, the observation of the longer period and the
details of the time-dependent density might be a valid source of information
on weak interactions among neutral particles.

The above approach can be easily generalized to the cases with
various interactions, and/or to the case of Fermion systems. It is
reasonable to expect that the above trapped scattering could be
experimentally realized via the progress of techniques in trapping
neutral particles by optical traps.

\begin{acknowledgments}
The support from the NSFC under the grant 10874249 is appreciated.
\end{acknowledgments}

\end{document}